# A SEM-NCA approach towards the impact of participative budgeting on budgetary slack and managerial performance: The mediating role of leadership style and leader-member exchange

**Corresponding author: PhD Candidate: Khalid Hasan Al Jasimee**
E-mail: khasan@correo.ugr.es
https://orcid.org/0000-0002-4425-5035
University of Al Qadisiyah[1], College of Biotechnology, Al Diwaniyah, Iraq
University of Granada[2], Faculty of Education, Economy, and Technology of Ceuta, Ceuta, Spain

**Co-author Associate Professor: Francisco Javier Blanco-Encomienda**
E-mail: jble@ugr.es
https://orcid.org/0000-0001-7449-3299
Supervisor University of Granada
Vice Dean for Research, Innovation, and Postgraduate Studies
Faculty of Education, Economy and Technology of Ceuta, Ceuta, (Spain)

**Abstract**

This study re-examines the impact of participative budgeting on managerial performance and budgetary slack, addressing gaps in current research. A revised conceptual model is developed, considering the conditioning roles of leadership style and leader-member exchange. The sample includes 408 employees with managerial experience in manufacturing companies. Hypotheses are tested using a combination of PLS-SEM and Necessary Condition Analysis (NCA). The results demonstrate that participative budgeting negatively affects budgetary slack and directly influences managerial performance, leadership style, and leader-member exchange. Moreover, leadership style and leader-member exchange moderate these relationships. The integration of NCA in management accounting research provides valuable insights for decision-makers, allowing for more effective measures. This study contributes by encouraging a complementary PLS-SEM and NCA approach to examine conditional effects. It also enhances understanding of budgetary control by highlighting the importance of leadership in influencing participative budgeting outcomes.

## 1. Introduction

It is increasingly important to understand management control systems, their tools, and how employees and managers interact with them (Chen et al., 2020). For example, budgetary control has been identified as one of the most effective managerial tools for saving costs, regulating organizations, and improving performance (Dos Santos et al., 2020).

Many successful companies, whether private or public, have implemented budgeting, like participative budgeting, to improve goal alignment between managers and subordinates (Bartech et al., 2022; Davila & Wouters, 2005).

The behavioral perspective suggests that budgets can be used as a mechanism to disseminate subordinates' motivation, plans, and objectives (Brink et al., 2018; Derfuss, 2016). Therefore, individuals' involvement in the budgetary strategy is essential to understand how the budget concerns individuals' cognitive states and behaviors at work, attitudes, and performance during budgeting processes (Brink et al., 2018; Degenhart et al., 2022).

Budgetary slack (BS) allows agents to reveal private information, which can result in inaccurate reporting and budgetary slack creation (Widanaputra & Mimba, 2014). On the other hand, if adequate compensation plans that reflect the manager's and agents' expectations are tied to goal attainment, participative budgeting contributes to more authentic budgets and improved managerial performance (PER) (Engelfried et al., 2021; Heinle et al., 2014).

The management accounting literature has yet to identify a conclusive direct relationship between participative budgeting (PB) and managerial performance (PER) (Derfuss, 2009, 2016) as well as participative budgeting (PB) and budgetary slack (BS) (De Baerdemaeker & Bruggeman, 2015; Van der Stede, 2000).

Prior studies have yielded inclusive results regarding the effects of participative budgeting (PB) on managerial performance (PER). Specifically, in both agency and psychological theory, participative budgeting positively affects managerial performance (Derfuss, 2016; Yao & Xiao-Na, 2018).

The prior literature indicates that participative budgeting has a positive relationship with performance (Chong & Chong, 2002; Chong et al., 2005; Dos Santos et al., 2020; Leach-López et al., 2009; Lunardi et al., 2019); and a weak or negative relationship (Cheng, 2012; Etemadi et al., 2009).

The inconclusive results reflect a lack of cognition of the mechanism by which participative budgeting (PB) influences managerial performance (PER). (Chong & Johnson, 2007). In addition, prior literature indicates that intervening factors could contribute to the explanation (Derfuss, 2016; Macinati et al., 2016).

Furthermore, understanding how vital variables affect budgetary slack (BS) is crucial since it is essential for many companies (Derfuss, 2012). Therefore, it is imperative to know how to manage budgetary slack (BS) to avoid adverse effects (Indjejikian & Matejka, 2006) and procure positive outcomes (Davila & Wouters, 2005). Several studies have found that participative budgeting (PB) affects budgetary slack (BS). However, prior research shows mixed evidence on how participative budgeting affects key outcomes like managerial performance and budgetary slack (Derfuss, 2016; Dunk, 1993). Heinle et al. (2014) suggest using agency and contingency models for inconsistent results between participative budgeting (PB) and budgetary slack (BS).

This research makes an important contribution by examining the mediating roles of leadership style (LS) and leader-member exchange (LMX) on the relationships between participative budgeting (PB) and managerial performance (PER)/budgetary slack (BS).

Prior studies have found mixed results on the direct PB-PER and PB-BS links, suggesting examining mediators can provide new insights. This responds to calls for research on mediators to clarify the mechanisms (Macinati et al. 2016).

LS and LMX are relevant mediators based on theory - LS affects subordinates' motivation and goals, while LMX impacts information sharing and relationships. Testing their mediating effects will reveal new insights on how leadership behaviors influence the effectiveness of PB.

Recent research also highlights the potential benefits of entrepreneurial bricolage for small tourism and hospitality firms. This involves creatively combining available resources to overcome limitations and quickly respond to changing conditions (Baker & Nelson, 2005).

For example, a study by Tajeddini et al. (2023) found that entrepreneurial bricolage can drive sustained competitive advantage for tourism SMEs by enabling differentiation and effective risk management despite resource constraints. This underscores the value of this flexible and adaptive approach for resilience in the turbulent tourism industry.

The ability to engage in entrepreneurial bricolage by recombining internal and external resources may be a valuable dynamic capability for tourism and hospitality SMEs to gain a competitive edge. However, research on its outcomes in this context remains limited.

A study by Tajeddini et al. (2020) found that committed frontline employees and effective leadership are key antecedents of service innovation in tourism firms. This underscores the importance of human-related factors like leadership in enabling innovation and performance in service contexts like tourism.

Leadership involves motivating employees, reducing disruptions during goal achievement, providing direction and support, and rewarding employees who achieve their goals (Champion-Hughes, 2001). Empirical findings regarding leadership style and participative budgeting are inconsistent due to previous studies that only examined the leader's perspective. A previous study hypothesized that leaders should treat their employees equally (Brink et al., 2018). Leaders' attitudes can affect employees. Since the LMX theory emerged, this position has changed.

LMX theory suggests that high-LMX supervisors establish effective connections with employees, just as LS develops constructive and stable relationships with all employees and colleagues (Yu & Liang, 2004).

Prior PB-PER literature has primarily addressed symmetrical statistical techniques, yielding a single predominant "net effects" framework. However, participative budgeting (PB) and managerial performance (PER) is a complex topic involving several factors interacting to enhance effectiveness. Hence, a study sensitive to complexity (such as NCA, detailed in Section 2) is especially relevant for its application in management accounting research.

Accordingly, budgeting professionals have advocated that future research investigate various combinations that may affect managerial performance (Derfuss, 2012, 2016).

Therefore, this study re-examines these relationships and makes two key contributions. First, it proposes a revised model that considers the conditioning roles of leadership style and leader-member exchange. Unlike past research focusing on direct effects, this study argues these leadership variables can change the impact of participative budgeting on outcomes. Second, it employs a complementary methodology combining PLS-SEM and Necessary Condition Analysis (NCA) to examine conditional relationships. This answer calls for greater attention to conditionality in management accounting research (Derfuss, 2012).

Assuming the rationales mentioned, two main aspects of the present study contribute to the field. First, this study establishes the relationship between participative budgeting (PB) and managerial performance (PER) as well as budgetary slack (BS). Our study proposes multiple mediator models to analyze the effect of participative budgeting on budgetary slack and managerial performance, despite extensive work on direct causality (see Figure 1).

Moreover, the prior literature emphasizes the importance of converting future studies to propose mediating role effects to assess the interfering relationships (Macinati et al., 2016). An invaluable contribution of this study is the formulation of a comprehensive framework that demonstrates that leadership style (LS) and leader-member exchange (LMX) mediate the influence of participative budgeting (PB) on managerial performance (PER) and budgetary slack (BS) and thus explains the mechanism by which each exogenous variable can specifically affect managerial performance.

Second, from a methodological perspective, an integrated technique is adopted to determine the relationships among the variables considered (Chong & Chong, 2002; Chong et al., 2005). Current research employs multiple non-symmetrical and symmetrical techniques. Necessary Condition Analysis (NCA) combines PLS-SEM outputs to pinpoint sufficient conditions in datasets to offer a more comprehensive understanding of the complex model design (Dul, 2016; Richter et al., 2020).

R-Studio and the NCA package were used to conduct the analysis. A recent study suggests that necessary and sufficient conditions play a critical role in management accounting analysis (Dul, 2016; Sukhov et al., 2022). PLS-predict was used to assess the accuracy of the proposed model on an out-of-sample basis (Shmueli et al., 2019).

Our research contributes to understanding computational dynamics based on non-symmetric relationships and their implications instead of exclusively focusing on symmetric relationships. It employs a combination of PLS-SEM, R-Studio, and the NCA package to identify the causal relationships among exogenous constructs and overall performance.

LS and LMX are better suited as mediators rather than antecedents/moderators. PB can influence a leader's style by making them more open and motivating. PB also improves LMX by enhancing information exchange and relationships between subordinates and supervisors. In turn, an open leadership style facilitates goal achievement and motivation from PB, while high-quality LMX allows better information sharing to improve decisions and performance. These mediate rather than moderate the effects of PB. The theories explain the mechanisms of how PB affects LS/LMX, which then impacts PER and BS.

The paper is designed as follows. Following this introduction, a literature review and hypotheses development are presented in Section 2, accompanied by a description of the research method in Section 3. Afterward, Section 4 summarizes statistical analyses and results obtained through PLS-SEM and Necessary Condition Analysis (NCA). Finally, our research discusses the principal conclusions derived from the study's results. Implications for managers are also provided to guide future decisions about leadership approaches to improve performance.

## 2. Conceptual model and hypotheses development
### 2.1. The relationship between PB and PER

Participative budgeting (PB) and managerial performance (PER) are extensively discussed in management accounting research, but the relationship is controversial and inconsistent (Derfuss, 2016). Participative budgeting (PB) is a mechanism in which subordinates set their budgets (Brink et al., 2018) and directly impact the budget (Dos Santos et al., 2020).

As a management tool, the budget facilitates three management activities: planning, coordinating, and controlling (Brink et al., 2018). Participative budgeting (PB) at different levels will improve the budget and influence individuals' mental effects, including increasing morale and self-efficacy (Brink et al., 2018; Derfuss, 2016; Hajdarowicz, 2022) as well as motivating individuals to accomplish budget targets (De Baerdemaeker & Bruggeman, 2015; Pamela, 2002).

Furthermore, participative budgeting (PB) generally influences organizational and individual performance, including attitudes, psychological well-being, and managerial performance (Kyj & Parker, 2008). managerial performance (PER) refers to managers' success in executing their management functions (Derfuss, 2016). Making decisions requires information when performing management functions. Therefore, management's ability to achieve truthful information is crucial to enhancing managerial performance (Aydiner et al., 2019).

Accordingly, participative budgeting (PB) affects managerial performance (PER) as measured by the achievement of management functions implemented by the budget. Also, participation in the budget will increase individual motivation, facilitating the achievement of budget targets and improving managerial performance (De Baerdemaeker & Bruggeman, 2015; Hajdarowicz, 2022; Pamela, 2002).

Prior literature (Chong et al., 2006; Covaleski et al., 2003; Her et al., 2019; Lau et al., 2018; Leach-López et al., 2009; Leach-López et al., 2007; Macinati et al., 2016; Maiga et al., 2014; Wagner et al., 2021) highlighted that the success of participative budgeting (PB) contributes directly to achieving budgetary goals and affects overall managerial performance. In light of the theory and previous studies, a hypothesis is formulated as follows:

**H1:** Participative budgeting has a positive effect on managerial performance.

### 2.2. The Influence of PB on BS

Budgetary slack (BS) is one of the most investigated issues in accounting management (Daumoser et al., 2018; De Baerdemaeker & Bruggeman, 2015; Dunk, 1993). Participative budgeting (PB) has been extensively researched and is assumed to impact budgetary slack (BS). Budgetary slack will be enhanced by a high level of participative budgeting (Young, 1985).

The participative budgeting process is evolving and allows the organizers of responsibilities to fill budgetary slack. Such events occur when the organizers of responsibilities are empowered to design their budget, which is used to assess their performance.

Managers are actively involved in composing their budgets under this budgeting system. Composing a budget is the first step in managing state finances.

In agency theory, Participative budgeting (PB) decreases information asymmetry by gaining personal information from agents. Budget proposals from lower managers will include personal information, making them more realistic. Subordinates will be more willing to help their superiors by providing their data, resulting in a more accurate budget proposal.

According to Hajdarowicz (2022), managers can boost their work morale and initiatives by participating in budgetary planning. Pamela (2002) defines work morale as satisfaction with one's job, superiors, and colleagues, while initiatives are ideas, opinions, and information planned. The morale created by such a condition would guide and enable workers and managers to formulate budgets.

Subordinate participation would enhance the sense of togetherness, belonging, and initiative. Thus, the decisions would be widely accepted. Moreover, participative budgeting may reduce conflicts between personal interests and organizational goals, improving performance among subordinates. participative budgeting may provide superiors with information about the present and future conditions of the environment (Daumoser et al., 2018). As a result, budgetary slack would be minimized, and budget proposals would be more accurate. A similar conclusion is made by Daumoser et al. (2018) and Maiga et al. (2014), who argue that participation tends to reduce budgetary slack.

Prior literature suggests that participative budgeting (PB) negatively affects budgetary slack (BS) (Davis et al., 2006; De Baerdemaeker & Bruggeman, 2015; Dunk, 1993). Subordinates would be encouraged to provide information to help the organization, thus decreasing budgetary slack (Daumoser et al., 2018; Maiga & Jacobs, 2007). Young (1985) indicates that participative budgeting (PB) positively impacts budgetary slack (BS). Based on the above and the contradiction of the conclusions of prior research, the following hypothesis is formulated:

**H2:** Participative budgeting has a negative effect on budgetary slack.

*2.3. The mediating role of leadership style on the relationship between PB and PER*

Leaders are transformation agents who strive to achieve equilibrium in their work activities so that their team members can achieve their goals (Lunardi et al., 2019). A leader usually has more influence and performs more leadership duties in a group (Kyj & Parker, 2008). According to Anderson and Sun (2017), scholars describe leadership as the power of persuasion that culminates in followers' loyalty.

Leadership theory contends that followers are affected by the leader's behavior, individually and collectively (Anderson & Sun, 2017). According to this theory, leadership is an activity that involves followers and leaders interacting. Leaders and subordinates establish high-level collaboration through exchanging esteem, confidence, and cooperation (Lumpkin & Achen, 2018).

As a result, the leadership style exhibited by the manager or leader stimulates and guides his employees to achieve higher performance and be accountable for budgeting. Thus, performance will improve through participative budgeting (PB) (Adler & Reid, 2008).

According to Kyj and Parker (2008), participation in budgetary decisions is directly related to the immediate superior's leadership style (LS). Baiocchi and Ganuza (2014) provide a framework for defining a subordinate's action using participative budgeting. This style may evoke certain types of human behavior. Therefore, organizations must promote organizational structures and leadership styles that encourage collaboration between superiors and subordinates to drive performance (Popli & Rizvi, 2016).

Literature suggests that leadership style (LS) is positively influenced by participative budgeting (PB) (Adler & Reid, 2008; Popli & Rizvi, 2016). However, they have no direct connection (Kyj & Parker, 2008; Lunardi et al., 2019). As a result, the following hypothesis is formulated:

**H3:** The positive relationship between participative budgeting and managerial performance will be moderated by leadership style, such that the relationship will be stronger when leaders exhibit a more participative/empowering leadership style, such as:

**H3a:** Leadership style moderates the relationship between participative budgeting and managerial performance.

    **H3b:** Leadership style positively affects managerial performance.

*2.4. Leadership style as a mediator of the relationship between PB and BS*

    Budget slack (BS) is an enduring control problem in numerous organizations that continues to receive widespread research interest. Economic and behavioral theories, such as principal agent, goal setting, or organizational fairness theory, have been incorporated to assess budget slack (BS), but the results vary depending on the variable considered (Derfuss, 2012).

    An individual's leadership style (LS) consists of integrity, self-awareness, courage, respect, empathy, and gratitude (Lumpkin & Achen, 2018). These leaders are regarded as public figures, and subordinates follow their behavior (Popli & Rizvi, 2016). It can occur since leadership style (LS) significantly impacts organizational policies (Lunardi et al., 2019). When leaders demonstrate the attributes above, they will eliminate budgetary slack (BS) by replicating their behavior (Daumoser et al., 2018).

    Budgetary slack (BS) represents the variance between actual and estimated budgets (Derfuss, 2012; Sheng, 2019). To improve managerial performance, subordinates should reduce the budget from the most accurate estimate.

    Additionally, subordinates engaged in budget preparation prefer to overestimate the budget from the most accurate estimate to receive recognition from leaders for meeting challenging budget objectives (Wagner et al., 2021).

    Chong and Johnson (2007) found that employees underestimate their capabilities when allowed to set their work expectations. In contrast to traditional budget control methods with relatively high expenditures, a flexible method such as the leadership style (LS) will decrease budgetary slack (Brink et al., 2018). Based on the above, the following hypothesis is proposed:

    **H4:** The relationship between participative budgeting and budgetary slack will be mediated by leadership style, in which participation interacts with leadership style, in turn decreasing budgetary slack, such as:

    **H4a:** Leadership style negatively affects budgetary slack.

*2.5. The mediating role of leader-member exchange on the relationship between PB and PER*

    Both psychological and agency theories suggest that participative budgeting (PB) positively impacts managerial performance (PER). According to agency theory, a leader assigns tasks to an efficient and flexible agent with superior expertise necessary for completing the task (Derfuss, 2016; Yao & Xiao-Na, 2018).

    Based on psychological theory, participative budgeting (PB) positively affects managerial performance (PER) via motivational and cognitive approaches (De Baerdemaeker & Bruggeman, 2015; Hajdarowicz, 2022; Pamela, 2002). By motivating methods that increase staff members' confidence, self-respect, and ego-engagement, participative budgeting improves performance by decreasing opposition to reform and growing approval of budget decisions (Derfuss, 2016).

    In participative budgeting, cognitive approaches are used to exchange information between superiors and subordinates, improving decision-making, including budgeting, leading to an increase in performance (Chong & Johnson, 2007).

    Psychological or agency theories do not suggest a direct association between participative budgeting (PB) and managerial performance (PER). Covaleski et al. (2003) described the relationship as dependent and mediated by other variables.

    In turn, the current study considers the role of leader-member exchange (LMX) in mediating the relationship between participative budgeting (PB) and managerial performance (PER). leader-member exchange (LMX) serves as a nexus between management trust and employee empowerment; leader-member exchange (LMX) provides intellectual stimulation, individualized consideration, motivational inspiration, and idealized influence, stimulating and increasing overall performance (Yu & Liang, 2004).

    Participative budgeting involves high managerial performance as one of its criteria and expectations

(Cheng, 2012). When employees interpret high leader-member exchange (LMX) as receiving incentives and advantages from their supervisors, they evaluate high managerial performance to meet supervisors' anticipations and conditions (Wang et al., 2019).

Employees strive for high task performance to reap incentives and other advantages (Yu & Liang, 2004). Providing employees with reciprocation may motivate them to reach high levels of managerial performance (PER) (Alfes et al., 2021). Employees are not motivated to invest energy when the leader-member exchange (LMX) is low-quality.

Leader-subordinate leader-member exchange (LMX) relationships are characterized by economic and social exchanges (Dulebohn et al., 2012). A high-quality relationship gives subordinates more inside information and increases their influence over decisions (Alfes et al., 2021).

Additionally, they receive higher incentive payments and pay increases. leader-member exchange (LMX) relationships have been positively linked to organizational citizenship behavior (Dulebohn et al., 2012) and innovative and creative behavior (Carnevale et al., 2017; Khalili, 2018; Mascareño et al., 2020), which are associated with individual performance. In high-quality leader-member exchange (LMX) relationships, subordinates perform well (Alfes et al., 2021; Dulebohn et al., 2012).

Estel et al. (2019) demonstrated that performance improved with a higher leader-member exchange (LMX) level (in-group) than with a lower leader-member exchange level (out-group). According to Stringer's (2006) research, leader-member exchange positively affects performance. When these two effects are combined, participative budgeting (PB) enhances managerial performance (PER) (Heinle et al., 2014).

High-quality LMX relationships are characterized by open communication, mutual trust, and reciprocal influence between a leader and follower (Dulebohn et al., 2012). Participative budgeting facilitates frequent communication and exchange of ideas between managers and subordinates. This should help develop higher-quality LMX relationships.

Employees who perceive high LMX view it as receiving benefits from their leader, and feel obligated to reciprocate with higher performance to meet the leader's expectations (Wang et al., 2019). Participative budgeting signals that a subordinate is trusted and valued by the leader. This can make the subordinate feel part of the "in-group" and motivate them to improve performance.

Subordinates in high LMX relationships receive more resources, information, and latitude from the leader (Carnevale et al., 2017). Participative budgeting provides a channel for subordinates to access information and influence decisions. This enhances what subordinates receive through a high LMX relationship.

High LMX relationships invoke a social exchange where the subordinate feels obligated to reciprocate the leader's support with higher effort and performance (Dulebohn et al., 2012). Participative budgeting can cultivate high-quality LMX where subordinates reciprocate with higher managerial performance. Accordingly, the following hypothesis is proposed:

**H5:** The relationship between participative budgeting and managerial performance will be mediated by leader-member exchange, in which participation interacts with a leader-member exchange, in turn increasing managerial performance, such as:

**H5a:** Participative budgeting positively affects leader-member exchange.
**H5b:** leader-member exchange positively affects managerial performance.

## 2.6. Leader-member exchange as a mediator of the relationship between PB and BS

Prior researchers presumed that leaders handled all their subordinates with identical attitudes. However, organizational leaders often display different attitudes toward their subordinates. Inconsistency in this attitude led to the development of the leader-member exchange theory (Kang et al., 2011; Wang et al., 2019).

Arikboga and Akdol (2017) propose that supervisors develop a close relationship with a small group of subordinates due to time constraints. Therefore, members of the group receive disproportionate support from the leadership. Trusted subordinates are grouped with leaders who give them special rights and unbalanced attention.

Conversely, subordinates who receive less time from their leaders have little control over their performance in giving awards (de Sousa Santos, 1998; Montambeault & Goirand, 2016).

A favorable special treatment can also increase employee contributions to the company. Leader-member exchange (LMX) is the basis for assessing leadership behavior in the organization, which results in a relationship between the leader and his subordinates, where subordinates become followers of the leader (Mascareño et al., 2020).

As Wang et al. (2019) argue, leader-member exchange (LMX) will improve the level of communication between supervisors and subordinates, thus improving their performance. Positive relationships will create feelings of volunteerism in employees to sacrifice for the company (Brink et al., 2018).

Budgetary slack (BS) is affected by internal and external influences. External aspects comprise participative budgeting (PB). Internal influences are derived from an individual's personality, such as a leader-member exchange (LMX).

As Sheng (2019) stated, human resources play a significant role in budget setting. The budget is influenced by the behavior of those closely engaged in budget setting. In light of this, the following hypothesis is proposed:

**H6:** The negative relationship between participative budgeting and budgetary slack will be moderated by leader-member exchange, such that the relationship will be stronger under conditions of high-quality LMX, such as:

**H6a:** Leader-member exchange moderates the relationship between participative budgeting and budgetary slack.

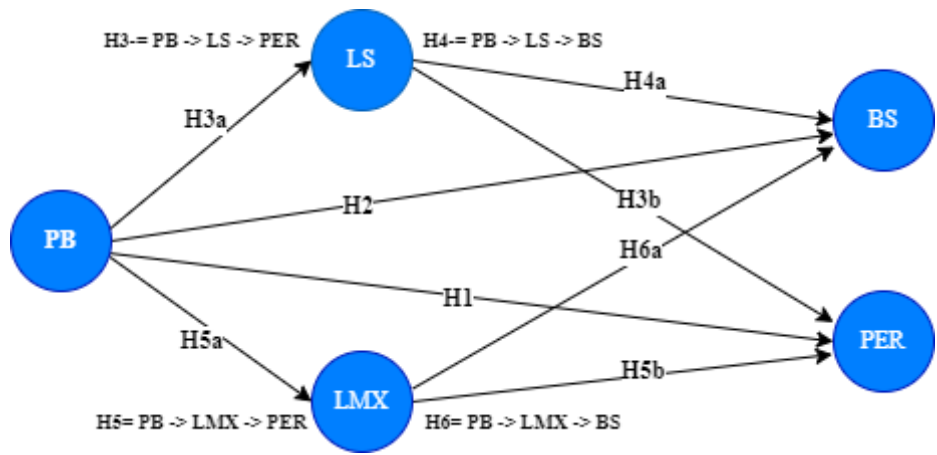

Fig. 1. Conceptual research model.

## 3. Method
### 3.1. Data collection

The target population consisted of employees of manufacturing companies in South Iraq. A questionnaire survey was conducted among mid and lower-level managers from 17 Iraqi units. We distributed 800 questionnaires, of which 456 were returned, resulting in a response rate of 57%. We were able to use 408 of these responses.

Therefore, the sample for this study included 408 employees with managerial experience from 17 manufacturing companies in Southern Iraq. Data were collected between November 2021 and February 2022.

The researchers distributed questionnaires to managers at various manufacturing firms to collect the data. Participating companies represented a range of industries including food and beverage, textiles, chemicals, metals, and machinery, and 56% of respondents held leadership positions and worked in the mentioned industries.

Responses were obtained from managers in various functions including manufacturing, product development, R&D, technology, and production. The average work experience of respondents was 10-20 years in managerial roles and 5 years in their current position with budgetary authority.

The survey included multiple items that covered each conceptual construct. Items were rated on a 7-point Likert scale from 1-strongly disagree to 7-strongly agree. The measures were adapted from previous scales. Thus, the six-factor PB was adapted from Chong et al. (2005), Parker and Kyj (2006), and Lunardi et al. (2019), and four-factor BS was adapted from Van der Stede (2000) and Kramer and Hartmann (2014).

Furthermore, an updated version of the "Leader Behaviour Description Questionnaire" was administered to assess LS, adapting a four-factor form from Kyj and Parker (2008), and Lunardi et al. (2019). LMX was evaluated based on the three-items scales proposed by Akdol Arikboga (2017)'.

In addition, PB was measured through a six-items scale adapted from Chong et al. (2005), Parker and Kyj (2006) and Lunardi et al. (2019). Finally, managers' performance was assessed according to an eight-factor scale adapted from Her et al. (2019).

*3.2. Symmetrical and asymmetrical modeling*

In order to determine the effects of multiple factors (BS, LS, LXM, PB), as well as the necessary and sufficient conditions for overall managerial performance, various analytical methods were used, including PLS-SEM, NCA-SEM, and PLS-predict. We analyzed the data according to the approaches presented by Richter et al. (2020).

The estimation of the conceptual framework was based on partial least squares structural equation analysis (PLS-SEM) (Cheah et al., 2021). It is recommended to implement PLS-SEM when dealing with complicated models, irregular data, and small samples (Shmueli et al., 2019).

PLS-path modeling is a practical approach for estimating complicated cause-and-effect models (Shmueli et al., 2019). In this regard, PLS-SEM is especially beneficial for developing and testing complex structural models early (Hair et al., 2016).

Through the demonstration of a relationship between exogenous and endogenous indicators, this study complements the necessary condition analysis (NCA) (Richter et al., 2020). NCA aims to determine the necessary conditions in data sets by applying innovative data analysis methods.

It is possible to identify necessary conditions by analyzing a scatter plot of endogenous and exogenous factors (Dul, 2016). This study used factor and composite scores as indicators to capture "unobservable, latent concepts" (Richter et al., 2020). The analysis was conducted using R-Studio and the NCA package.

Finally, PLS-predict was used to generate in-out-of-sample predictions to eliminate predictable out-of-sample states. Shmueli et al. (2016) proposed incorporating PLS-SEM into PLS-predict to achieve accurate predictions.

In contrast to standard structural model evaluation metrics such as R-Square and Q-Square, PLS-predict measures a model's ability to forecast new outcomes from the sample. PLS-predict applies the k-fold validation procedure from predictive analysis to PLS-SEM (Shmueli et al., 2019). Evaluation of the study is based on the model's ability to make falsifiable predictions.

## 4. Results
### 4.1. Measurement validation

Before testing the hypotheses, we conducted a confirmatory factor analysis to analyze the construction's validity and reliability using the PLS-SEM application (Table 1). As a result, all variables in the research have factor loadings above 0.50 (Hair et al., 2021). Furthermore, Cronbach's alpha, Rho-A, and Rho-C exceeded a recommended threshold of 0.700, confirming construct reliability (Ebrahimi et al., 2021; Hair et al., 2021; Nekmahmud et al., 2022).

AVE greater than 0.5 also demonstrated the convergence validity of the measurement model (Hair et al., 2021; Shmueli et al., 2019).

**Table 1. Validity and reliability of the constructs.**

| Construct/Item | Loading | VIF | Alpha | Rh-A | Rh-C | AVE |
|---|---|---|---|---|---|---|
| **Participatory Budget (PB)** | | | | | | |
| PB1 | .753 | 1.608 | .831 | .833 | .877 | .543 |
| PB2 | .775 | 1.809 | | | | |
| PB3 | .725 | 1.654 | | | | |
| PB4 | .728 | 1.774 | | | | |
| PB5 | .751 | 1.771 | | | | |
| PB6 | .686 | 1.411 | | | | |
| **Managerial Performance (PER)** | | | | | | |
| PER1 | .760 | 1.975 | .914 | .915 | .930 | .624 |
| PER2 | .770 | 2.007 | | | | |
| PER3 | .792 | 2.186 | | | | |
| PER4 | .812 | 2.226 | | | | |
| PER5 | .817 | 2.465 | | | | |
| PER6 | .797 | 2.293 | | | | |
| PER7 | .779 | 2.202 | | | | |
| PER8 | .791 | 2.179 | | | | |
| **Budgetary Slack (BS)** | | | | | | |
| BS1 | .863 | 2.264 | .879 | .880 | .917 | .733 |
| BS2 | .848 | 2.21 | | | | |
| BS3 | .853 | 2.208 | | | | |
| BS4 | .861 | 2.28 | | | | |
| **Leadership Style (LS)** | | | | | | |
| LS1 | .848 | 2.038 | .833 | .849 | .889 | .668 |
| LS2 | .837 | 1.949 | | | | |
| LS3 | .868 | 2.263 | | | | |
| LS4 | .705 | 1.452 | | | | |
| **Leadership Member Exchange (LMX)** | | | | | | |
| LMX1 | .708 | 1.38 | .767 | .825 | .862 | .677 |
| LMX2 | .748 | 1.051 | | | | |
| LMX3 | .702 | 1.362 | | | | |

Discriminant validity was determined considering the Fornell-Larcker criterion (Fornell & Larcker, 1981). As shown in Table 2, the square root of the AVE (bold numbers) is greater than the correlations among constructs. Thus, discriminant validity was established.

**Table 2. Discriminant validity of measures.**

| Construct | BS | LS | LMX | PB | PER |
|---|---|---|---|---|---|
| BS | **.856** | .823 | .793 | .750 | .722 |
| LS | .714 | **.817** | .849 | .767 | .711 |
| LMX | .579 | .641 | **.719** | .718 | .697 |
| PB | .643 | .642 | .525 | **.737** | .793 |
| PER | .648 | .624 | .528 | .695 | **.790** |

**Note:** BS, budgetary slack; LS, leadership style; LMX, leader–member exchange; PB, participative budgeting; PER, managerial performance.

### 4.2. Structural equation model

Using the PLS algorithm and pairing deletion to account for the absence of data, the structural equation model could be assessed after a valid and reliable measurement model has been developed. VIF values were lower than 5 (see Table 1), indicating that multicollinearity is not a concern (Ebrahimi et al., 2021). On the other hand, a value below three indicates immaculate conditions (Hair et al., 2021; Shmueli et al., 2019).

R-Square, Q-Square, and the significance of paths were used to evaluate the structural equation model. R-Square is calculated for each structural path for the endogenous constructs to determine the model's goodness. R-Square should be equal to or greater than 0.1 (Hay & McCabe, 2002). As a result, all R-square values are above 0.1, thus proving the model's predictive capability (Table 3).

**Table 3. Q-Square and R-square.**

| Construct | Q-Square | RMSE | MAE | R-square | R-square adjusted |
|---|---|---|---|---|---|
| BS | .406 | .776 | .573 | .580 | .577 |
| LS | .406 | .776 | .599 | .412 | .411 |
| LXM | .267 | .862 | .685 | .275 | .273 |
| PER | .477 | .729 | .534 | .569 | .565 |
| SRMR= 0.857; NFI= 0.828; d-G= 0.466; d-ULS= 1.753 | | | | | |

**Note:** BS, budgetary slack; LS, leadership style; LMX, leader–member exchange; PB, participative budgeting; PER, managerial performance.

Moreover, the Q-square provides evidence that endogenous constructs are predictive. Predictive relevance can be determined by a Q-square above 0 (Dul, 2016). The results show that the constructs can be accurately predicted.

Finally, to avoid model misspecification, the standardized root means square residual (SRMR) serves as a goodness-of-fit parameter (Hair et al., 2021). The model fit was evaluated using SRMR, produced by calculating correlation matrices from the sample covariance matrix and predicted covariance matrix. Based on PLS-SEM results, 0.857 is considered a satisfactory fit (Shmueli et al., 2019).

To estimate the statistical significance of the path coefficients, 10,000 bootstrapped data sets were used with bias-and-error-corrected path coefficient estimates. Based on a one-tailed test at 0.05, confidence intervals that differ from zero indicate significant relationships (see Table 4).

#### 4.2.1. Direct effects analysis

Based on the results shown in Table 4 and Figure 2, all direct hypotheses in the research were supported at a significance level of .05. In particular, PB significantly positively impacts PER (β = .409, t = 8.317, p = .000) and negatively affect BS (β = -.285, t = -5.409, p = .000), thus supporting H1 and H2. Moreover, H3a, H3b, and H4a are supported as PB directly impacts LS (β = .642, t = 17.467, p = .000), as well as LS positively, impacts PER (β = .136, t = 2.314, p = .001) and BS (β = .434, t = 7.067, p = .000). Besides, PB significantly affected LMX (β = 0.525, t = 12.662, p = .000), and LMX was found to have a significant effect on PER (β = .089, t = 2.028, p = .021) and BS (β = .151, t = 3.212, p = .001), respectively, thereby supporting H5a, H5b, and H6a.

**Table 4. Direct relationship testing.**

| Construct | Path coefficient | SD | t value (bootstrap) | P values | CI | |
|---|---|---|---|---|---|---|
| | | | | | Lower | Upper |
| H1 -> PB -> PER | .409 | .049 | 8.317 | .000 | .333 | .495 |
| H2 -> PB -> BS | -.285 | .053 | -5.409 | .000 | .198 | .372 |
| H3a -> PB -> LS | .642 | .037 | 17.467 | .000 | .581 | .702 |
| H3b-> LS -> PER | .136 | .059 | 2.314 | .001 | .038 | .230 |
| H4a -> LS -> BS | .434 | .061 | 7.067 | .000 | .329 | .532 |
| H5a -> PB -> LMX | .525 | .041 | 12.662 | .000 | .458 | .594 |
| H5b-> LMX -> PER | .089 | .044 | 2.028 | .021 | .018 | .162 |
| H6a -> LMX -> BS | .151 | .047 | 3.212 | .001 | .074 | .229 |

**Note:** BS, budgetary slack; LS, leadership style; LMX, leader–member exchange; PB, participative budgeting; PER, managerial performance.

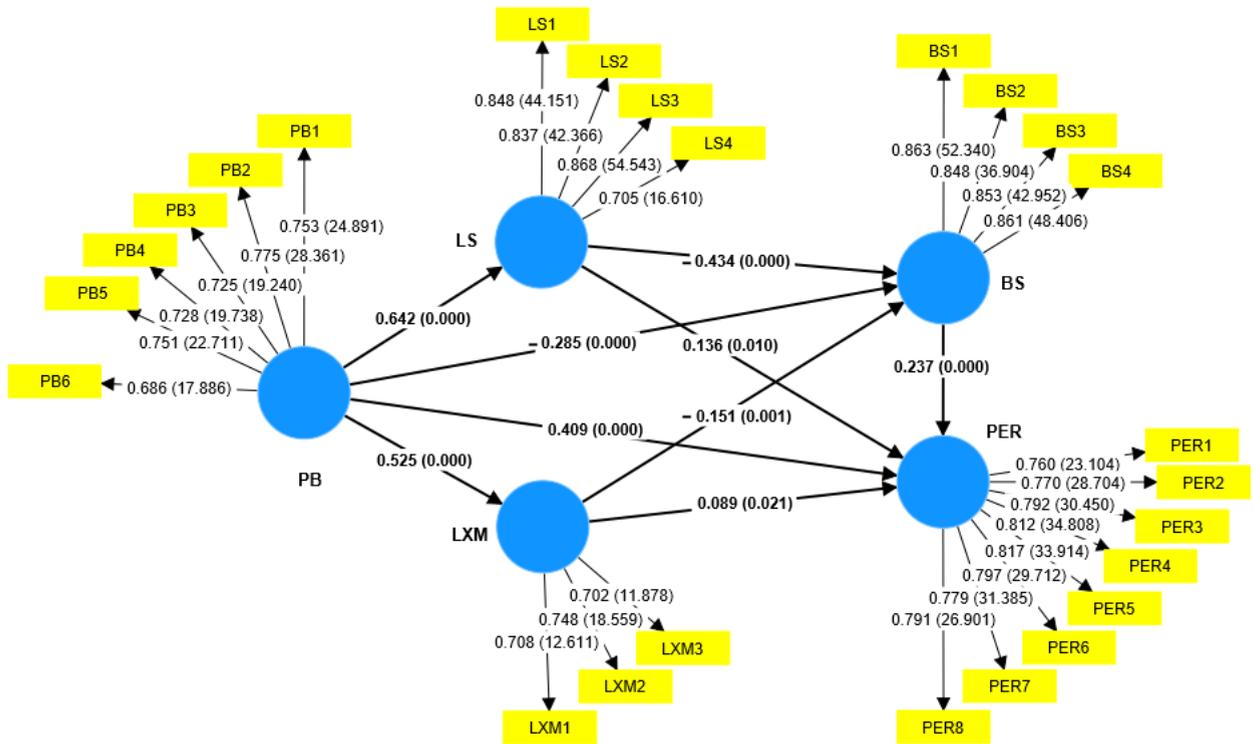

Fig. 2. Structural equation model estimated.

### 4.2.2. Mediating effects analysis

LS and LMX were assessed for their mediating role through mediation analysis. Based on the results (see Table 5), the relationship between PB and PER has a complementary (partial mediation) of LS ($\beta$ = .087, t = 2.285, p = .011) and LMX ($\beta$ = .047, t = 1.930, p = .027), providing support for H3 and H5. Moreover, the results indicate that the relationship between PB and BS has a complementary (partial mediation) of LS ($\beta$ = -.279, t = -6.641, p = .000) and LMX ($\beta$ = -.079, t = -3.050, p = 0.001). Therefore, H4 and H6 are supported.

**Table 5. Mediation effects analysis.**

| Construct | Path coefficient | SD | t value (bootstrap) | P values | CI | |
|---|---|---|---|---|---|---|
| | | | | | Lower | Upper |
| **H3**-> PB -> LS -> PER | | .087 | .038 | 2.285 | .011 | .024 | .149 |
| **H4**-> PB -> LS -> BS | | -.279 | .042 | -6.641 | .000 | .209 | .347 |
| **H5**-> PB -> LMX -> PER | | .047 | .024 | 1.930 | .027 | .009 | .089 |
| **H6**-> PB -> LMX -> BS | | -.079 | .026 | -3.050 | .001 | .038 | .123 |

**Note:** BS, budgetary slack; LS, leadership style; LMX, leader–member exchange; PB, participative budgeting; PER, managerial performance.

### 4.3. Necessary Condition Analysis (NCA)

PLS-SEM was enhanced with Necessary Condition Analysis (NCA) to assess the relationship between BS, LS, LXM, PB, and overall managerial performance. NCA was conducted with latent-variable scores derived from PLS-SEM for exogenous constructs and overall managerial performance. R software was used to import these scores and run the NCA package. A non-decreasing step function, the ceiling envelope-free disposal hull, was used to plot predictors and outcomes. Discrete data also explained the CE-FDH ceiling line, albeit with limited ranges (Dul et al., 2021).

We could determine to what extent a given exogenous factor restricts overall managerial performance by separating the data set containing observations from the data set without any observed data. Similarly, a ceiling line indicates the minimum exogenous constructs required to achieve PER. Figure 3 illustrates the data's midpoint by an OLS regression curve.

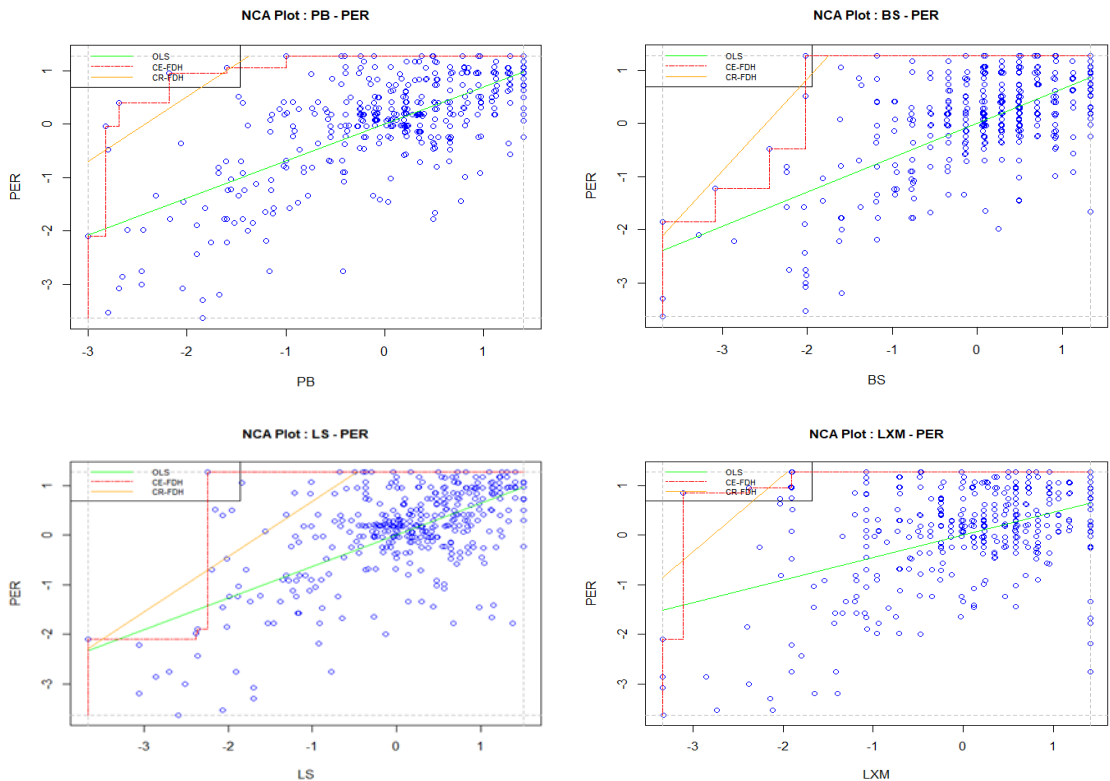

Fig. 3. Scatter plot of the predictor variables

A random sample size of 10,000 is recommended for testing latent variable effect sizes ($d$) (Dul, 2016; Dul et al., 2020, 2021). Thus, $d$ varies from 0 to 1. Specifically, Dul (2016) identified that "$0 < d < 0.1$ as a 'small effect,' $0.1 \leq d < 0.3$ as a 'medium effect,' $0.3 \leq d < 0.5$ as a 'large effect,' and $d \geq 0.5$ as a 'very large effect.'"

Based on the NCA results, BS ($\beta$ = .173, P = .000), LS ($\beta$ = 0.187, P = .000), LMX ($\beta$ = .053, P = .001), and PB (H4: $\beta$ = 0.071, P = .000) are sufficient conditions for the overall PER. Table 6 indicates that BS and LS (0.1 < d < 0.3) moderate overall PER. Whereas LMX and PB (0 ≤ d ≤ 1) have a "small effect" (Dul, 2016).

**Table 6. Celling line effect size CE-FDH.**

|     | Path coefficients | 95.00% | Permutation p-value |
| --- | --- | --- | --- |
| BS  | .173 | .056 | .000 |
| LS  | .187 | .093 | .000 |
| LMX | .053 | .04  | .010 |
| PB  | .071 | .041 | .000 |

**Note:** BS, budgetary slack; LS, leadership style; LMX, leader–member exchange; PB, participative budgeting.

For further clarification, a bottleneck analysis was conducted. Table 7 outlines the values required for each outcome measure (PER). BS, LS, LMX, and PB must be at least 4.412%, 3.431%, 1.225%, and 3.431%, respectively, to ensure an acceptable level of PER (40–90%). The endogenous rate of PER (100%) should be at least 4.412%, 3.431%, 4.657%, and 17.402%, respectively.

If exogenous constructs are not achieved, PER will not be high. Exogenous constructs are required to improve the overall PER. For PER to manifest, a specific level of exogenous material is required (4.412%, 3.431%, 4.657%).

**Table 7. Bottelenck CE-FDH percentile.**

| Bottleneck | PER | BS | LS | LXM | PB |
| --- | --- | --- | --- | --- | --- |
| 0.00% | -3.637 | 0.0 | 0.0 | 0.0 | 0.0 |
| 10% | -3.146 | 0.0 | 0.0 | 0.0 | 0.0 |
| 20% | -2.654 | 0.0 | 0.0 | 0.0 | 0.0 |
| 30% | -2.163 | 0.0 | 0.0 | 0.0 | 0.0 |
| 40% | -1.672 | 1.225 | 3.431 | 1.225 | 0.245 |
| 50% | -1.180 | 1.716 | 3.431 | 1.225 | 0.245 |
| 60% | -0.689 | 1.716 | 3.431 | 1.225 | 0.245 |
| 70% | -0.198 | 4.412 | 3.431 | 1.225 | 0.245 |
| 80% | 0.293 | 4.412 | 3.431 | 1.225 | 1.471 |
| 90% | 0.785 | 4.412 | 3.431 | 1.225 | 3.431 |
| 100% | 1.276 | 4.412 | 3.431 | 4.657 | 17.402 |

**Note:** BS, budgetary slack; LS, leadership style; LMX, leader–member exchange; PB, participative budgeting; PER, managerial performance.

### 4.4. Assessing the model's predictive power PLS-predict

Finally, PLS-predict was applied to estimate the model's predictive power out of the sample. The MECM's generalizability can be established by evaluating its accuracy in predicting the outcome value of novel cases using the PLS-predict tool (Shmueli et al., 2019).

PLS path models were used to estimate LM benchmark values by linear regression for a given endogenous construct's indicator versus a preconfigured exogenous construct's indicator (Marin-Garcia & Alfalla-Luque, 2019).

Based on Shmueli et al. (2019), we compared Root Mean Square Error (RMSE) with naive linear model (LM) values. The results indicated that all $Q^2_{predict}$ values of the PLS-SEM were above 0 (see Table 8).

Consequently, PLS-SEM yields smaller RMSE values than LM benchmarks. According to this, RMSE provides high predictive ability (Shmueli et al., 2019).

**Table 8. Model's predictive power PLS-predict for PER.**

|  | Q2predict | PLS_SEM RMSE | LM_RMSE | ∆RMSE |
|---|---|---|---|---|
| BS1 | 0.314 | 1.152 | 1.161 | -0.009 |
| BS2 | 0.253 | 1.166 | 1.174 | -0.008 |
| BS3 | 0.300 | 1.192 | 1.215 | -0.023 |
| BS4 | 0.318 | 1.170 | 1.183 | -0.013 |
| LS1 | 0.304 | 1.152 | 1.166 | -0.014 |
| LS2 | 0.284 | 1.133 | 1.148 | -0.015 |
| LS3 | 0.314 | 1.118 | 1.133 | -0.015 |
| LS4 | 0.175 | 1.293 | 1.299 | -0.006 |
| LMX1 | 0.071 | 1.370 | 1.384 | -0.014 |
| LMX2 | 0.234 | 1.257 | 1.262 | -0.005 |
| LMX3 | 0.070 | 1.414 | 1.415 | -0.001 |
| PER1 | 0.261 | 1.182 | 1.184 | -0.002 |
| PER2 | 0.243 | 1.261 | 1.269 | -0.008 |
| PER3 | 0.282 | 1.231 | 1.235 | -0.004 |
| PER4 | 0.345 | 1.178 | 1.194 | -0.016 |
| PER5 | 0.333 | 1.194 | 1.208 | -0.014 |
| PER6 | 0.308 | 1.226 | 1.228 | -0.002 |
| PER7 | 0.305 | 1.231 | 1.233 | -0.002 |
| PER8 | 0.284 | 1.229 | 1.236 | -0.007 |

**Note:** BS, budgetary slack; LS, leadership style; LMX, leader–member exchange; PB, participatory budget; PER, managerial performance.

## 5. DISCUSSION

*5.1. Conclusions*

This study intends to explain how participatory budgeting (PB) affects managerial performance (PER) and budgetary slack (BS). Additionally, the current study considers the role of the leadership approach in mediating the relationship between PB-PER and PB-BS to develop an integrated and comprehensive research model based on a contingency approach (Pradana, 2021).

Data analysis results indicate that all hypotheses posed in this study are supported and statistically significant. The conclusions of this study support several theories of agency and psychology. However, they remain compatible with the regulations of a patriarchal society, such as Iraq. In this budgeting situation, leadership, agency, and psychological theories can all be used to explain the superior's behavior.

Therefore, researchers can define budgeting procedures more comprehensively and validly using multiple approaches (Covaleski et al., 2003). The following points provide additional details on our model analysis.

PB-PER is reported to increase when PB is implemented. PB affects PER positively. As a result, PB can improve managers' decision-making ability and positively determine strategies to impact PER in a manufacturing company. The budgeting process will increase each manager's accountability for their current responsibilities. Consequently, all employees, especially middle-level managers, are expected to be active in constructing budgets to enhance managerial performance.

As lower-level managers are better acquainted with the natural conditions of their parts, they are more likely to propose budgets, thereby helping managers achieve budget targets jointly formulated and set by the company. Thus, PB will increase PER. There is a positive correlation between PB and PER.

Thus, the findings are consistent with those found in previous studies (Chong et al., 2006; Covaleski et al., 2003; Her et al., 2019; Lau et al., 2018; Leach-López et al., 2009; Leach-López et al., 2007; Macinati et al., 2016; Maiga et al., 2014; Parker & Kyj, 2006; Wagner et al., 2021).

Statistically, study results indicate a direct relationship between PB and BS, as evidenced by their negative influence on budgetary slack in the companies studied. In addition, participation may provide superiors with information regarding the environment's current and future requirements. Thus, budgetary slack would be reduced, and budget proposals would be more reliable. Prior literature reached similar conclusions (Davis et al., 2006; De Baerdemaeker & Bruggeman, 2015; Dunk, 1993; Daumoser et al., 2018; Maiga & Jacobs, 2007).

Motivating and managing employees to achieve their highest performance level is a critical aspect of companies. Creating effective strategic decisions and facilitating change is how a leader creates change. Thus, leadership involves more than managing daily functions and designing a system that guides performance. Overall, empirical results support the theoretical framework. Developing more comprehensive models may better understand PB's role in PER.

Regarding the contingency approach, the results reveal that leaders with a considerate leadership style tend to encourage employee participation in budgeting. Furthermore, the leadership style of manufacturing firms was partially significant and mediated the relationship between PB and PER. Moreover, the present study found that superiors allow employees to participate when budget targets are used to estimate subordinates.

LS can demonstrate leadership traits such as integrity, self-awareness, courage, respect, empathy, gratitude, and respect for others to reduce company budget slack. Since these leaders are considered authoritative by their subordinates, they follow their behavior. Therefore, LS partially influences BS and mediates the relationship between PB and BS. In addition, BS measures the discrepancy between actual and estimated budgets (Daumoser et al., 2018). Therefore, reducing BS can improve the performance of subordinates.

According to the results, performance improved with a higher LMX level (in the group) compared to a lower LMX level (outside the group). According to Wang's (2016) research, LMX positively affects performance. As a result of the combination of these two effects, PB increases PER (Heinle et al., 2014). We concluded from our analysis that LMX is partially influenced by PB and mediates the relationship between PB and PER.

Additionally, the study found that LMX mediates the relationship between PB and BS and partially significantly affects BS. As a result of the study, external and internal factors have been found to influence budget slack. Budget participation is an external factor. The internal factors are determined by the individual's personality (LMX). Budgets are influenced by the behavior of those involved in their preparation. Communication between leadership and subordinates could improve budget preparation. This study found that improving communication between leadership and subordinates can improve subordinates' involvement in budget preparation and reduce budgetary slack.

This paper makes several notable contributions to previous management accounting research published in Middle Eastern journals. First, it develops and tests an integrated model examining mediating mechanisms between participative budgeting and outcomes. Most prior Middle East studies focus just on direct relationships.

Second, it combines PLS-SEM and Necessary Condition Analysis methodologies to provide a more nuanced understanding. The application of NCA in management accounting is still rare, especially in Middle Eastern research.

Third, the data comes from Iraqi companies, providing valuable insights from this understudied context. Moreover, it explores Western-developed management accounting concepts within the patriarchal context of Iraq and the Middle East region, assessing their applicability.

Lastly, this study provides information on planning, controlling, and making the right decisions to compete, develop, and enhance managerial performance in Southern Iraqi companies. A manager's strategy for participatory budgeting is critical, especially regarding managers' performance.

We developed a common approach to identify bottlenecks and clarify relationships between predictors and outcomes (PLS-SEM and NCA). Latent-variable scores were derived from PLS-SEM for exogenous constructs and overall managerial performance. R software was used to import these scores and run the NCA package. The ceiling line CE-FDH measures were applied to compare predictors and outcomes. Discrete data also explained the CE-FDH ceiling line but at relatively low levels (Dul et al., 2021).

*5.2. Implications*

The results of this paper have relevant implications for scholars and managers. From an academic perspective, it contributes to understanding the PB-BS relationship with LS and LMX as mediating factors.

The study confirms the role of PB in using the leadership approach to improve PER. Through taking on PB activities, the organization can improve performance-level results by facilitating the leadership approach through the LS and LMX roles. This will, in turn, result in setting and reducing BS. To enhance managerial performance, organizations should focus their efforts on PB initiatives that could provide information on planning, controlling, and making the right decisions.

Furthermore, using NCA results, managers can combine PB and BS dimensions with leadership approaches LS and LMX dimensions to fulfill the prerequisites (i.e., causality), resulting in higher PER. The logic of necessity identifies conditions that cause high performance, whereas the logic of sufficiency identifies minimum requirements. PLS-SEM is an emerging methodology in management accounting research despite the sufficiency logic's prevalence. NCA still needs to be explored in management accounting research.

NCA and PLS-SEM are complementary approaches so researchers can apply them to develop further explanations. By focusing on the necessary and sufficient factors, decision-makers can take more effective measures. Our contribution encourages and guides researchers to adopt a complementary approach, combining PLS-SEM and NCA to explore lower and upper levels.

Finally, using a sample of manufacturing companies in southern Iraq, this study evaluated a conceptual framework primarily constructed from Western management theories. Particularly in the case of high task uncertainty, the model was well supported. Comparing this model's validity across different cultural samples is an interesting issue. We can pinpoint the most relevant characteristics of a culture that may not be relevant in another, enhancing our understanding of this critical organizational management tool.

*5.3. Limitations and further research*

The limitations of the present study also represent avenues for future research. This study represents a novel approach among the few earlier studies to combine symmetrical and non-symmetrical methods. This approach can also be applied to future studies, advancing the research area. In addition, it would be beneficial to test the model in different cultural contexts to strengthen its adaptability. There are some limitations to this study.

A sample of top budget officials in Iraq was selected. Future studies may benefit from a broader sample. The study concentrated on four factors; additional factors could be considered in future studies (e.g., ethical leadership style, moral identity, and servant leadership style) to evaluate the effect of participative budgeting on managerial performance and establish configurations that may cause organizational outcomes by using these causal recipes.

Furthermore, future research could examine the links between budget participation and budgetary slack in light of moral identity and ethical leadership style as mediating factors. Accordingly, servant leadership and ethical leadership styles may differ in their impact on budgetary slack depending on whether budget targets operate as fundamental control mechanisms versus uncertain environments requiring further research.

**Conflict of interest:** All authors have no conflicts of interest.

# References

Adler, R. W., & Reid, J. (2008). The effects of leadership styles and budget participation on job satisfaction and job performance. *Asia-Pacific Management Accounting Journal*, *3*(1), 21–46.
Akdol, B., & Arikboga, F. S. (2017). Leader-member exchange as a mediator of the relationship between servant leadership and job satisfaction: A research on Turkish ICT companies. *International Journal of Organizational Leadership*, *6*, 525–535.
Alfes, K., Veld, M., & Fürstenberg, N. (2021). The relationship between perceived high-performance work systems, combinations of human resource well-being and human resource performance attributions and engagement. *Human Resource Management Journal*, *31*(3), 729–752.
Anderson, M. H., & Sun, P. Y. T. (2017). Reviewing leadership styles: Overlaps and the need for a new 'full-range theory. *International Journal of Management Reviews*, *19*(1), 76–96.
Aydiner, A. S., Tatoglu, E., Bayraktar, E., & Zaim, S. (2019). Information system capabilities and firm performance: Opening the black box through decision-making, performance, and business-process performance. *International Journal of Information Management*, *47*, 168–182.
Baiocchi, G., & Ganuza, E. (2014). Participatory budgeting as if emancipation mattered. *Politics & Society*, *42*(1), 29–50.
Beer, M. (2003). Why total quality management programs do not persist: the role of management quality and implications for leading a TQM transformation. *Decision Sciences*, *34*(4), 623–642.
Biswas, S. R., Uddin, M. A., Bhattacharjee, S., Dey, M., & Rana, T. (2022). Ecocentric leadership and voluntary environmental behavior for promoting sustainability strategy: The role of psychological green climate. *Business Strategy and the Environment*, *31*(4), 1705–1718.
Brink, A. G., Coats, J. C., & Rankin, F. W. (2018). Who 's the boss? The economic and behavioral implications of various characterizations of the superior in participative budgeting research. *Journal of Accounting Literature*, *41(1)*, 89–105.
Carnevale, J. B., Huang, L., Crede, M., Harms, P., & Uhl-Bien, M. (2017). Leading to stimulate employees' ideas: A quantitative review of leader-member exchange, employee voice, creativity, and innovative behavior. *Applied Psychology*, *66*(4), 517–552.
Champion-Hughes, R. (2001). Totally integrated employee benefits. *Public Personnel Management*, *30*(3), 287–302.
Cheah, J. H., Roldán, J. L., Ciavolino, E., Ting, H., & Ramayah, T. (2021). Sampling weight adjustments in partial least squares structural equation modeling: Guidelines and illustrations. *Total Quality Management & Business Excellence*, *32*(13–14), 1594–1613.
Chen, C. X., Lill, J. B., & Vance, T. W. (2020). Management control system design and employees' autonomous motivation. *Journal of Management Accounting Research*, *32*(3), 71–91.
Cheng, M. T. (2012). The joint effect of budgetary participation and broad-scope management accounting systems on management performance. *Asian Review of Accounting*, *20*(3), 184–197.
Chong, V. K., & Johnson, D. M. (2007). Testing a model of the antecedents and consequences of budgetary participation on job performance. *Accounting and Business Research*, *37*(1), 3–19.
Chong, V. K., Eggleton, I. R. C., & Leong, M. K. C. (2005). The impact of market competition and budgetary participation on performance and job satisfaction: a research note. *The British Accounting Review*, *37*(1), 115–133.
Chong, V. K., Eggleton, I. R. C., & Leong, M. K. C. (2006). The multiple roles of participative budgeting on job performance. *Advances in Accounting*, *22*, 67–95.
Chong, V., & Chong, K. M. (2002). The role of feedback on the relationship between budgetary participation and performance. *Pacific Accounting Review*, *14*(2), 33–55.
Covaleski, M. A., Evans, J. H., Luft, J. L., & Shields, M. D. (2003). Budgeting research: Three theoretical perspectives and criteria for selective integration. *Journal of Management Accounting Research*, *15*, 3–50.
Daumoser, C., Hirsch, B., & Sohn, M. (2018). Honesty in budgeting: a review of morality and control aspects in the budgetary slack literature. *Journal of Management Control*, *29*(2), 115–159.
Davila, T., & Wouters, M. (2005). Managing budget emphasis through the explicit design of conditional budgetary slack. *Accounting, Organizations and Society*, *30*(7–8), 587–608.
Davis, S., DeZoort, F. T., & Kopp, L. S. (2006). The effect of obedience pressure and perceived responsibility on management accountants' creation of budgetary slack. *Behavioral Research in Accounting*, *18*(1), 19–35.
De Baerdemaeker, J., & Bruggeman, W. (2015). The impact of participation in strategic planning on managers' creation of budgetary slack: The mediating role of autonomous motivation and affective organisational commitment. *Management Accounting Research*, *29*, 1–12.
de Sousa Santos, B. (1998). Participatory budgeting in Porto Alegre: toward a redistributive democracy. Politics & Society, 26(4),

461–510.

Degenhart, L., Zonatto, V. C. D. S., & Lavarda, C. E. F. (2022). Effects of psychological capital and managerial attitudes on the relationship between budgetary participation and performance. *Revista Contabilidade & Finanças*, *33*, 216-231.

Derfuss, K. (2009). The relationship of budgetary participation and reliance on accounting performance measures with individual-level consequent variables: a meta-analysis. *European Accounting Review*, *18*(2), 203–239.

Derfuss, K. (2012). *Budget slack: Some meta-analytic evidence*. 8th Conference on New Directions in Management Accounting, Brussels, Belgium.

Derfuss, K. (2016). Reconsidering the participative budgeting–performance relation: A meta-analysis regarding the impact of level of analysis, sample selection, measurement, and industry influences. *The British Accounting Review*, *48*(1), 17–37.

Dos Santos, G. C., Grespan, C. A., & Gaio, L. E. (2020). Health service in Brazilian private and public hospitals: Budgetary participation, feedback, and performance from clinical managers' perception. *African Journal of Business Management*, *14*(11), 457–466.

Dul, J. (2016). Necessary condition analysis (NCA) logic and methodology of "necessary but not sufficient" causality. *Organizational Research Methods*, *19*(1), 10–52.

Dul, J., Van der Laan, E., & Kuik, R. (2020). A statistical significance test for necessary condition analysis. *Organizational Research Methods*, *23*(2), 385–395.

Dul, J., Vis, B., & Goertz, G. (2021). Necessary Condition Analysis (NCA) does exactly what it should do when applied properly: a reply to a comment on NCA. *Sociological Methods & Research*, *50*(2), 926–936.

Dulebohn, J. H., Bommer, W. H., Liden, R. C., Brouer, R. L., & Ferris, G. R. (2012). A meta-analysis of antecedents and consequences of leader-member exchange: Integrating the past with an eye toward the future. *Journal of Management*, *38*(6), 1715–1759.

Dunk, A. S. (1993). The effect of budget emphasis and information asymmetry on the relation between budgetary participation and slack. *Accounting Review*, 400–410.

Ebrahimi, P., Khajeheian, D., & Fekete-Farkas, M. (2021). A SEM-NCA approach towards social networks marketing: Evaluating consumers' sustainable purchase behavior with the moderating role of eco-friendly attitude. *International Journal of Environmental Research and Public Health*, *18*(24), 13276.

Engelfried, T., Hirdina, L., Rauschenberger, F., & Schulz, L. (2021). What we know and what we don't know about participative budgeting: a systematic literature review. *Management Studies*, *11*(2), 11–20.

Estel, V., Schulte, E. M., Spurk, D., & Kauffeld, S. (2019). LMX differentiation is good for some and bad for others: A multilevel analysis of effects of LMX differentiation in innovation teams. *Cogent Psychology*, *6*(1), 1614306.

Etemadi, H., Dilami, Z. D., Bazaz, M. S., & Parameswaran, R. (2009). Culture, management accounting and managerial performance: Focus Iran. *Advances in Accounting*, *25*(2), 216–225.

Fornell, C., & Larcker, D. (1981). Evaluating structural equation models with unobserved variables and measurement error. *Journal of Marketing Research*, *18*, 39–50.

Hair, J. F., Astrachan, C. B., Moisescu, O. I., Radomir, L., Sarstedt, M., Vaithilingam, S., & Ringle, C. M. (2021). Executing and interpreting applications of PLS-SEM: Updates for family business researchers. *Journal of Family Business Strategy*, *12*(3), 100392.

Hajdarowicz, I. (2022). Does participation empower? The example of women involved in participatory budgeting in Medellin. *Journal of Urban Affairs*, *44*(1), 22–37.

Hay, L. E., & McCabe, G. J. (2002). Spatial variability in water-balance model performance in the conterminous United States. *Journal of the American Water Resources Association*, *38*(3), 847–860.

Heinle, M. S., Ross, N., & Saouma, R. E. (2014). A theory of participative budgeting. *The Accounting Review*, *89*(3), 1025–1050.

Her, Y. W., Shin, H., & Pae, S. (2019). A multigroup SEM analysis of moderating role of task uncertainty on budgetary participation-performance relationship: Evidence from Korea. *Asia Pacific Management Review*, *24*(2), 140–153.

Indjejikian, R. J., & Matejka, M. (2006). Organizational slack in decentralized firms: The role of business unit controllers. *The Accounting Review*, *81*(4), 849–872.

Jarrar, N. S., & Smith, M. (2014). Innovation in entrepreneurial organizations: A platform for contemporary management change and a value creator. *The British Accounting Review*, *46*(1), 60–76.

Kang, D. S., Stewart, J., & Kim, H. (2011). The effects of perceived external prestige, ethical organizational climate, and leader-member exchange (LMX) quality on employees' commitments and subsequent attitudes. *Personnel Review*, *40*(6), 761-784.

Khalili, A. (2018). Creativity and innovation through LMX and personal initiative. *Journal of Organizational Change Management*, *31*(2), 323–333.

Kramer, S., & Hartmann, F. (2014). How top-down and bottom-up budgeting affect budget slack and performance through social and economic exchange. *Abacus*, *50*(3), 314–340.

Kyj, L., & Parker, R. J. (2008). Antecedents of budget participation: Leadership style, information asymmetry, and evaluative use of budget. *Abacus: A Journal of Accounting, Finance, and Business Studies*, *44*(4), 423–442.

Lau, C. M., Scully, G., & Lee, A. (2018). The effects of organizational politics on employee motivations to participate in target setting and employee budgetary participation. *Journal of Business Research*, *90*, 247–259.

Leach-López, M. A., Stammerjohan, W. W., & Lee, K. S. (2009). Budget participation and job performance of South Korean

managers mediated by job satisfaction and job-relevant information. *Management Research Review*, *32*(3), 220–238.

Leach-López, M. A., Stammerjohan, W. W., & McNair, F. M. (2007). Differences in the role of job-relevant information in the budget participation-performance relationship among US and Mexican managers: A question of culture or communication. *Journal of Management Accounting Research*, *19*(1), 105–136.

Lumpkin, A., & Achen, R. M. (2018). Explicating the synergies of self-determination theory, ethical leadership, servant leadership, and emotional intelligence. *Journal of Leadership Studies*, *12*(1), 6–20.

Lunardi, M. A., Costa da Silva Zonatto, V., & Constâncio-Nascimento, J. (2019). Relationship between leadership style, encouragement of budgetary participation, and budgetary participation. *Estudios Gerenciales*, *35*(150), 27–37.

Macinati, M. S., Bozzi, S., & Rizzo, M. G. (2016). Budgetary participation and performance: The mediating effects of medical managers' job engagement and self-efficacy. *Health Policy*, *120*(9), 1017–1028.

Maiga, A. S., & Jacobs, F. A. (2007). The moderating effect of manager's ethical judgment on the relationship between budget participation and budget slack. *Advances in Accounting*, *23*, 113–145.

Maiga, A. S., Nilsson, A., & Jacobs, F. A. (2014). Assessing the impact of budgetary participation on budgetary outcomes: The role of information technology for enhanced communication and activity-based costing. *Journal of Management Control*, *25*(1), 5–32.

Marin-Garcia, J., & Alfalla-Luque, R. (2019). Key issues on Partial Least Squares (PLS) in operations management research: A guide to submissions. *Journal of Industrial Engineering and Management*, *12*(2), 219–240.

Mascareño, J., Rietzschel, E., & Wisse, B. (2020). Leader-Member Exchange (LMX) and innovation: A test of competing hypotheses. *Creativity and Innovation Management*, *29*(3), 495–511.

Montambeault, F., & Goirand, C. (2016). Between collective action and individual appropriation: The informal dimensions of participatory budgeting in Recife, Brazil. Politics & Society, 44(1), 143–171.

Nekmahmud, M., Ramkissoon, H., & Fekete-Farkas, M. (2022). Green purchase and sustainable consumption: A comparative study between European and non-European tourists. *Tourism Management Perspectives*, *43*, 100980.

Pamela, R. (2002). A critical evaluation of the effect of participation in budget target setting on motivation. *Managerial Auditing Journal*, *17*(3), 122–129.

Parker, R. J., & Kyj, L. (2006). Vertical information sharing in the budgeting process. *Accounting, Organizations and Society*, *31*(1), 27–45.

Popli, S., & Rizvi, I. A. (2016). Drivers of employee engagement: The role of leadership style. *Global Business Review*, *17*(4), 965–979.

Pradana, B. G. V. (2021). The role of psychological capital and leader member-exchange on participatory budgeting and managerial performance. *Media Ekonomi Dan Manajemen*, *36*(1), 11–26

Richter, N. F., Schubring, S., Hauff, S., Ringle, C. M. & Sarstedt, M. (2020). When predictors of outcomes are necessary: Guidelines for the combined use of PLS-SEM and NCA. *Industrial Management & Data Systems, 120*(12), 2243–2267.

Sheng, S. (2019). *Literature review on the budget slack*. 3rd International Conference on Education, Management Science and Economics, Singapore.

Shmueli, G., Ray, S., Estrada, J. M. V., & Chatla, S. B. (2016). The elephant in the room: Predictive performance of PLS models. *Journal of Business Research*, *69*(10), 4552–4564.

Shmueli, G., Sarstedt, M., Hair, J. F., Cheah, J. H., Ting, H., Vaithilingam, S., & Ringle, C. M. (2019). Predictive model assessment in PLS-SEM: Guidelines for using PLSpredict. *European Journal of Marketing*, *53*(11), 2322–2347.

Stringer, L. (2006). The link between the quality of the supervisor-employee relationship and the level of the employee's job satisfaction. *Public Organization Review*, *6*(2), 125–142.

Sukhov, A., Olsson, L. E., & Friman, M. (2022). Necessary and sufficient conditions for attractive public Transport: Combined use of PLS-SEM and NCA. *Transportation Research Part A: Policy and Practice*, *158*, 239–250.

Tajeddini, K., Gamage, T. C., Tajeddini, O., & Kallmuenzer, A. (2023). How entrepreneurial bricolage drives sustained competitive advantage of tourism and hospitality SMEs: The mediating role of differentiation and risk management. *International Journal of Hospitality Management*, *111*, 103480.

Tajeddini, K., Martin, E., & Altinay, L. (2020). The importance of human-related factors on service innovation and performance. *International Journal of Hospitality Management*, *85*, 102431.

Tekleab, A. G., Reagan, P. M., Do, B., Levi, A., & Lichtman, C. (2021). Translating corporate social responsibility into action: A social learning perspective. *Journal of Business Ethics*, *171*(4), 741–756.

Van der Stede, W. A. (2000). The relationship between two consequences of budgetary controls: Budgetary slack creation and managerial short-term orientation. *Accounting, Organizations and Society*, *25*(6), 609–622.

Wagner, J., Petera, P., Popesko, B., Novák, P., & Šafr, K. (2021). Usefulness of the budget: The mediating effect of participative budgeting and budget-based evaluation and rewarding. *Baltic Journal of Management*, *16*(4), 602–620.

Wang, Z., Liu, Y., & Liu, S. (2019). Authoritarian leadership and task performance: The effects of leader-member exchange and dependence on leader. *Frontiers of Business Research in China*, *13*(1), 1–15.

Widanaputra, A. A., & Mimba, N. (2014). The influence of participative budgeting on budgetary slack in composing local governments' budget in Bali province. *Procedia-Social and Behavioral Sciences*, *164*, 391–396.

Yao, G. E., & Xiao-Na, D. (2018). *Research on the impact of budgetary participation on management performance based on meta-*

*analysis*. 5th International Conference on Management Science and Management Innovation, Wuhan, China.

Young, S.M. 1985. Participative Budgeting: The Effects of Risk Aversion and Assymetric Informations on Budgetary Slack. Journal of Accounting Research. 23 (2): 829-842.

Yu, D., & Liang, J. (2004). A new model for examining the leader-member exchange (LMX) theory. *Human Resource Development International*, *7*(2), 251–264.